# Combining Multiple View Components for Exploratory Visualization

Vladimir Guchev  
University of Turin, Italy

Paolo Buono[†]  
University of Bari, Italy

Cristina Gena[‡]  
University of Turin, Italy

**ABSTRACT**

The analysis of structured complex data, such as clustered graph based datasets, usually applies a variety of visual representation techniques and formats. The majority of currently available tools and approaches to exploratory visualization are built on integrated schemes for simultaneous displaying of multiple aspects of studying objects and processes. Usually, such schemes partition screen space that is composed of multiple views and adopt interaction patterns to focus on data-driven items. Widely known concepts as overview plus-detail and focus-plus-context are ambiguous in interpretation by means of technical terms. Therefore, their implementation by UI design practitioners need reviews and a classification of the basic approaches to visual composition of graphical representation modules. We propose a description of basic components of the view and focus and an overview of their multiple combinations.

Index Terms: K.5.1 [Human-centered computing]: Visualization—Visualization systems and tools Visualization toolkits; K.5.2 [Human-centered computing]: Visualization—Visualization techniques Graph drawings;

## 1 ELEMENTS OF VIEW AND FOCUS

The interactive visual exploration of complex multidimensional datasets, due to the variety of tasks and related techniques, usually assumes the different data manipulations at different levels of representation. Therefore, the single-view implementation is often not enough for full-fledged analytical work. The demand for data visualization tools based on multiple views that represent relations among dataset through multiple simultaneous representations is increasing [1, 11]. The research directions in this area includes: the organization of static predefined screen spaces for multiple dashboard-like views [4], the development of dynamic user-controlled screen spaces [8], the compressed in-screen and offscreen summaries at viewport borders [3], the use of whitespace and semi-transparent watermarks to enrich the presentation by providing an auxiliary context for selected data points.

The modern frameworks for data analysis allow flexible data manipulations through multi-layered data layout. The flexibility during the exploration is especially important for graph-based representations characterized by an intrinsic complex structure that must be visualized. Node-link representations are the most obvious way to visualize relationships but contain issues related to their readability, and require a trade-off between the level of detail of shown graph elements at different zoom levels and their layout on the screen. The exploration assumes selective extraction of content from a set of visual layers and elements of data layouts, according to the user choices. A data layout can be considered as a layer with different properties, such as detail of data points, spatial or structural embedding, level of visual aggregation, semantic annotation, distribution in space and time. Thus, the view available to the user reflects a part of data layout in the context of exploration.

---
[†]paolo.buono@uniba.it  
[‡]cristina.gena@unito.it

This work proposes an approach to form the *base*, *focus* and *context* views, built according to the scheme shown in Fig. 1 (left). The *base view* reflects the scope of dynamic query result from the data layout, which changes during the user exploration: specifically, during zoom and panning [2].

The *focus* consists of the focus view on relevant items and is built on a data selection inside a focal area. The *context view* represents the projection of the base view to adjusted general components of the data layout. The main parts of the focus component can be used both in conjunction or independently. The independent use of focal area marking usually applied for the attention management (e.g. system-initiated highlighting of a found data element) or as a visual feedback for user-performed manipulation over a data element; the independent use of focused view elements may be implemented for the fixed scope or structure of related data content.

In the case of a focus component, the captured focal area may have an additional display in order to augment (by annotation or extension), amplify (to change the presentation form) or compress (to reduce the content form) the visual representation.

The geometric combinations of the focal area and focused view elements are shown in Fig. 1 (center). When focal area and focused views are bound tightly (Fig. 1, center, top row), the following patterns are available: both elements are equal in area and position (center: a) —as applied in various semantic lenses (e.g. magic, "see-through", "bring neighbours") [5,15]; the smaller focal area is placed below bigger focused view (center: b) —the part of base view is hidden by the overlaying layer or displayed via partial grid deformation or local layout re-arrangement [9] around focal area, as it is done in fish-eye lens or tablelens [3]; the adjacent combination of both focal area and focused view is a single bundled unit in the form of floating viewframe (center: c). Such solution gives a better understanding of the scale proportions and is used as the detail-on-demand.

When focal area and focused view are independent (Fig. 1, center, bottom row), the following patterns are available: the focused view is a fixed nested/built-in frame inside a base view (center: d) that shows the data inside the focal area —this approach is widely applied in geospatial systems [15]; the focused view is a fixed or draggable adjacent external to a base view frame (center: e) bound with the scope of focal area —such composition is used in tabletop display i-Loupe [17]. Differently by these example, Fig. 1 (center, f) implements the concept of a reducing glass that shows the navigation context of the base view in a global layout scope.

On the screen space use (Fig. 1, right), the information capacity of a view component is extendable by viewport frame areas, and juxtaposition of both, data-driven and supplementary layers, through whitespace (to mark graph visualization elements) or semitransparent watermark-style screens (quick look or pop-up annotations). The space around viewframes may fit pointers to indicate selections, as well as can be suitable for contextual awareness during exploration through off-screen summaries [7] and marginal diagrams [12].

## 2 COMBINING MULTIPLE COMPONENTS

Data visualization based on multiple views opens wide opportunities for simultaneous analytical presentation of multiple data sources and their various properties, including type, dimension, scale, level of details, format, semantic encoding, and degree of abstraction. The implementation of different interactive patterns for multiple views and foci offers the possibilities for intuitive and flexible userperformed data manipulations such as exploration, searching, aggregation and summarization, coordinated selection, filtering, creation, and annotation.

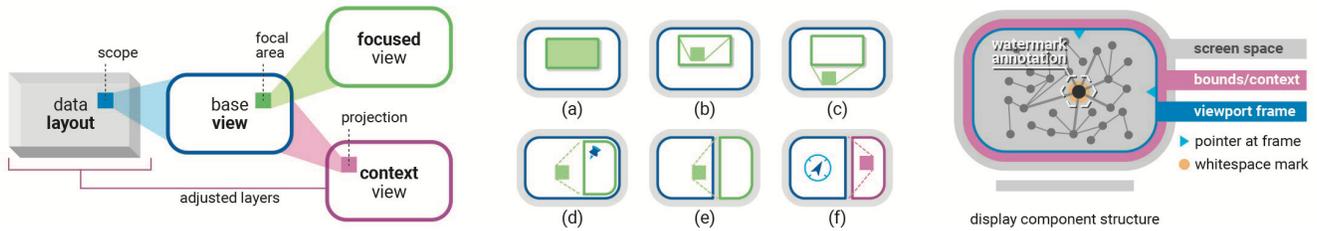

Figure 1: The scheme (left) of forming focus/context views for selected data scope. The ways of coupled representation (center) of the focal area and a focused or context view. The main parts (right) of the view component and basic marking techniques of the selected element.

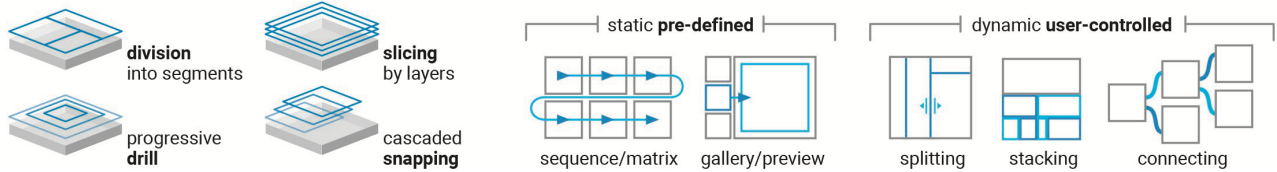

Figure 2: Schemes of approaches to sequential data extraction (left) for structured multiple view representations (right).

The basic strategies of sequential data extraction for multiple view representation (Fig. 2, left) can be reduced to the following types: segmentation of content for subsequent comparison or matching the resulting parts; layered slicing for selective representing different levels of data layout; stepwise drilling along with gradual increasing or decreasing of the data scope for progressive detailing; sequential snapping of the uniform in size and structure captures for obtaining of consistently fragmented view sets.

The main ways of positioning the multiple views (Fig. 2, right) are static, pre-defined, formed on the basis of screen design pattern [19] and dynamic user-controlled, formed in accordance with the rules of visual organization [8,11]. The basic approaches to the interactive coupling of different views are one-sided or both-sided coordination and synchronization [18]. The visual coupling of different views is achievable via adjacent placement, use of common coordinated space, and implicit anchoring of views content [10,16]

In the combination of view and focus, the use of one-one or onemany, or many-many correspondences is relevant. Examples for each combination are the following: single focal area and single focused view: zoom lens in interactive scatterplot [14]; single focal area and multiple focused views: linked multi-form views of spacetime cube visualization [13]; multiple focal areas and single focused view: ensemble visualization [6]; multiple focal areas and multiple focused views: MultiLens [9].